\documentclass[12pt]{iopart}
\pdfoutput=1

\usepackage[colorlinks=true,linkcolor=black, urlcolor = black, citecolor = blue, anchorcolor = blue]{hyperref}

\usepackage{graphicx}
\usepackage[tight]{subfigure}
\usepackage{multirow}
\usepackage{subeqnarray}
\usepackage{wasysym}
\usepackage{eufrak}
\usepackage{amsopn}

\graphicspath{{figs/}}

\usepackage{eulervm}
\usepackage{wasysym}

\usepackage[british]{babel}

\usepackage{amssymb}

\usepackage{framed}

\usepackage{vmargin}

\usepackage{epsfig}

\usepackage{color}

\usepackage{alltt}

\usepackage{pstricks,pst-node,pst-tree,pst-grad,pst-text,graphics}

\usepackage{dcolumn}

\usepackage{url}

\usepackage{cases}

\usepackage[square,numbers]{natbib}
\usepackage{ifthen}

\usepackage{afterpage}

\usepackage{comment}

\usepackage{supertabular}

\usepackage{xspace}

\usepackage{setspace}

\usepackage[bf,footnotesize]{caption}

\usepackage{textcomp}

\usepackage{tikz}
\usetikzlibrary{decorations.pathmorphing}
\usetikzlibrary{decorations.markings}
\usetikzlibrary{arrows,shapes,snakes,automata,backgrounds,petri}

\newcommand{\gpcomment}[1]{[GP: {\it #1}]}
\renewcommand{\gpcomment}[1]{}

\setpapersize{A4}
\newlength{\bibmarkkeyAleft}

\newlength{\bibmarkkeyBleft}

\newlength{\bibmarkkeyCleft}

\newlength{\bibmarkkeyDleft}

\newcommand{\gpvec}[1]{\mathbf{#1}}
\newcommand{\kvec}{\gpvec{k}}
\newcommand{\xvec}{\gpvec{x}}
\newcommand{\yvec}{\gpvec{y}}

\newcommand{\Adim}{\mathtt{A}}
\newcommand{\Bdim}{\mathtt{B}}

\newcommand{\Ldim}{\mathtt{L}}
\newcommand\newblock{\hskip .11em\@plus.33em\@minus.07em}

\newcommand{\renorm}{\mathcal{R}}

\newcommand{\Ps}{P_{\textnormal{\scriptsize s}}}

\usetikzlibrary{calc}
\usetikzlibrary{patterns}
\usetikzlibrary{decorations.text}
\tikzset{
xxtsubstrate/.style={decorate, 
line width=1pt,
draw=olive, 
decoration=snake, 
segment amplitude=0.75mm, 
line after snake=0.25mm,
line before snake=0.25mm
},
tsubstrate/.style={decorate, 
line width=1pt,
draw=olive, 
decoration=snake, 
segment amplitude=0.5mm, 
segment length=5pt,
segment amplitude=0.2mm, 
line after snake=1mm,
line before snake=1mm
},
Bsubstrate/.style={decorate, 
line width=1pt,
draw=orange, 
decoration=snake,
segment length=5pt,
segment aspect=0,
segment amplitude=0.5mm, 
line after snake=0mm,
line before snake=0mm
},
substrate/.style={decorate, 
line width=1pt,
draw=orange,
decoration=snake, 
segment length=5pt,
segment amplitude=0.5mm, 
line after snake=0.5mm,
line before snake=0.5mm
},
activity/.style={very thick,draw=red,postaction={decorate},
decoration={markings,mark=at position .5 with
{\arrow[draw=red]{>}}}},
tactivity/.style={thick,draw=red,postaction={decorate},
decoration={markings,mark=at position .5 with
{\arrow[draw=red]{>}}}},
tEPSactivity/.style={thick,draw=red,postaction={decorate},
decoration={markings,mark=at position .55 with
{\arrow[draw=red]{>}}}},
tAactivity/.style={thick,draw=red},
Aactivity/.style={very thick,draw=cyan},
tSactivity/.style={thick,draw=red,postaction={decorate},
decoration={markings,mark=at position .7 with
{\arrow[draw=red]{>}}}},
Sactivity/.style={very thick,draw=red,postaction={decorate},
decoration={markings,mark=at position .7 with
{\arrow[draw=red]{>}}}},
polarity/.style={decorate, 
line width=1pt,
draw=red,
decoration={markings,mark=between positions 0 and 1 step 1.1mm with {\draw[red,thick]  (0,0) circle (0.03)} },
segment length=5pt,
segment amplitude=0.5mm, 
},
Bpolarity/.style={decorate, 
line width=1.5pt,
draw=red,
segment length=5pt,
segment amplitude=0.5mm, 
},
density/.style={ 
line width=1.5pt,
draw=black,
densely dashed,
segment length=5pt,
segment amplitude=0.5mm, 
},
}

\newcommand{\ave}[1]{\left\langle #1 \right\rangle}
\newcommand{\Psurv}{\Ps}

\newcommand{\plaind}{\mathrm{d}}
\newcommand{\plaint}{\mathrm{t}}

\newcommand{\tildephi}{\widetilde\phi}

\newcommand{\tildepsi}{\widetilde{\psi}}
\newcommand{\tilderho}{\widetilde{\rho}}
\newcommand{\tildenu}{\widetilde{\nu}}
\newcommand{\tildeU}{\widetilde{\mathcal{U}}}

\newcommand{\Eqref}[1]{Eq.~(\ref{eq:#1})}
\newcommand{\Esref}[1]{Eqs.~(\ref{eq:#1})}

\newcommand{\elabel}[1]{\label{eq:#1}}
\newcommand{\deltabar}{\delta\mkern-8mu\mathchar'26}
\newcommand{\imag}{\mathring{\imath}}
\newcommand{\dbar}{\plaind\mkern-6mu\mathchar'26}

\renewcommand{\exp}[1]{\mathchoice{e^{#1}}{\operatorname{exp}\!\left(#1\right)}{\operatorname{exp}\!\left(#1\right)}{\operatorname{exp}\!\left(#1\right)}}
\newcommand{\corresponding}{\hat{=}}

\newcommand{\mA}{\mathcal{A}}
\newcommand{\mD}{\mathcal{D}}

\newcommand{\mJ}{\mathcal{J}}
\newcommand{\mK}{\mathcal{K}}
\newcommand{\mO}{\mathcal{O}}
\newcommand{\mU}{\mathcal{U}}

\usepackage{xspace}
\newcommand{\latin}[1]{{\it #1}}
\newcommand{\ie}{\latin{i.e.}\@\xspace}

\newcommand{\Erefs}[1]{Eqs.~(\ref{eq:#1})}

\usepackage{textcomp}

\begin{document}
 
\title{The Concealed Voter Model is in the Voter Model universality class}

\author{Rosalba Garcia-Millan}
\ead{garciamillan16@imperial.ac.uk}
\address{Department of Mathematics, Imperial College London, London SW7 2AZ, United Kingdom}
\address{Centre for Complexity Science, Imperial College London, London SW7 2AZ, UK}

\date{\today}

\begin{abstract}
The Concealed Voter Model (CVM) is an extension of the original Voter Model, where two different opinions compete until consensus is reached. In the CVM, agents express an opinion not only publicly, they also hold a private opinion, which they may disclose or change. In this paper I derive the critical exponents of both models via renormalised field theory and show that both belong to the compact directed percolation universality class.
\end{abstract}



\section{Introduction \label{intro}}
The Concealed Voter Model (CVM) is a simple agent-based model  of opinion formation in social
networks that takes into account the consistency or the lack thereof
 between a voter's private and publicly expressed opinions \cite{Gastner_2018,Gastner_2019}.
In the original Voter Model (VM), agents adopt the expressed opinion of one of their neighbours at random
 until the system reaches consensus \cite{Liggett:1985}. Modelling the private opinion of agents in the CVM
  is introduced as an extension to the VM in \cite{Gastner_2018}.

The VM is a special case of the Domany-Kinzel cellular automaton, which is in the universality class of
compact directed percolation, also known as the VM universality class \cite{DomanyKinzel:1984,Dickman:1995,
Dornic:2001}.
To determine whether the CVM belongs to the same universality class, I calculate the critical exponents characterising
 the spreading of an opinion from a localised source via a path integral approach.
 In \cite{Howard:1998}, a path-integral method was used to compute the persistence probability on a slight variation of
 the VM.

\begin{figure}
\centering
\subfigure[]{ \label{VM01} \includegraphics[]{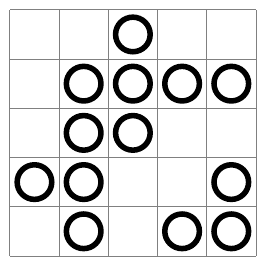}}
\subfigure[]{ \label{VM04} \includegraphics[]{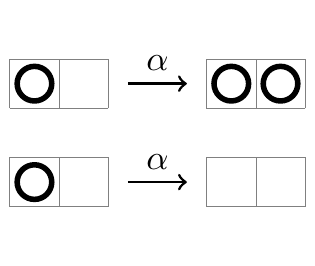}}
\captionsetup{justification=raggedright,singlelinecheck=false}
\caption{\label{VM} The voter model (opinions are represented by circles and empty spaces).
\subref{VM01} Sample of a particle configuration in the VM in $d=2$.
\subref{VM04} The process of copying at the cluster boundary.} 
\end{figure}

In the VM, each agent expresses one of two opinions, which may be thought of as particles and empty sites on a 
$d$-dimensional lattice, Fig.~\ref{VM}.
At rate $\alpha>0$, every agent \textbf{copies} the opinion expressed by one of their randomly
chosen neighbours. The system is initialised with a single particle, and the process of copying is repeated
until the system reaches one of the two absorbing states: all sites either empty or occupied.
In the VM, activity only happens at the boundaries between clusters of empty and occupied sites.
The critical point of the VM separates the two phases 
where, in the thermodynamic limit, particles are subject to extinction with probability $1$
and where there is a positive probability of indefinite survival.
Microscopically, the VM is at criticality if given a pair of neighbours at the interface between clusters $A+\emptyset$,
the two possible outcomes after copying, $\emptyset+\emptyset$ and $A+A$, are equally likely. 
If, for instance, the former is more likely, then, on average, there is a net loss of particles that
gives rise to the mass in the field theory.
The VM at criticality is also known as unbiased voter \cite{Dickman:1995} and it is representative of so-called 
neutral theories \cite{martinello2017neutral}.

Observables commonly considered to characterise such critical processes are \cite{GrassbergerdelaTorre:1979,
Munoz:1997}: the expected number of particles $n(t)$ at time $t$,
the mean square spread of particles $R^2(t)$ (or radius of gyration),
 and  the survival probability $\Ps(t)$, namely, the probability that there is at least one particle in the system.
At criticality, these three observables have an asymptotic algebraic scaling in time,
\begin{equation}
\ave{n(t)} \propto t^\eta, \elabel{crit_exp} \quad 
R^2(t) \propto t^z,  \quad 
\Psurv(t) \propto t^{-\delta}, 
\end{equation}
whose exponents $\eta$, $z$ and $\delta$ characterise the VM universality class.
The critical exponent $z$ is closely related to the dynamical critical exponent $\hat{z}$, 
defined as $\tau_t\propto \xi^{\hat{z}}$, where $\tau_t$ and $\xi$ are temporal and spatial correlation lengths,
via the relation   $z=2/\hat{z}$ \cite{Grassberger:1995}. 

Due to the duality between the VM and coalescent random walks \cite{Dickman:1995,
Dornic:2001, Munoz:1997}, it is known that $\eta=0$ and $z=1$ in all dimensions,
 and $\delta=1/2$ at dimension $d=1$ and $\delta=1$ for dimensions $d\geq2$,
satisfying the hyperscaling relation \cite{Dickman:1995}
\begin{equation}
\delta+\eta=\frac{1}{2}dz 
\end{equation}
in dimensions $d\leq2$.

In Sec.~\ref{FT_VM}, I derive the field theory of the VM and calculate the critical exponents;
 in Sec.~\ref{FT_CVM}, I incorporate the dynamics of the CVM in the field theory and calculate the critical
exponents. Finally, I discuss the results in Sec.~\ref{Conclus}.

\section{Critical exponents of the Voter Model \label{FT_VM}}
The critical exponents of the VM are well understood \cite{Dickman:1995, Munoz:1997}.
Nevertheless, in this section, the exponents of the VM are derived via a path integral approach
  to lay the 
 foundations of the derivation of the critical exponents of the CVM in Sec.~\ref{FT_CVM}.
 
 The configuration of the system is given by the set $\{m\}$ of occupation numbers $m_\xvec$,
  which is the number of particles at lattice site $\xvec$.
The dynamics of the VM are condensed in the following master equation,
\begin{eqnarray}
\fl\dot{P} (\{m\};t) =
 \sum_{\xvec,\yvec} \Big\{ 
\alpha \, (m_\xvec+1)\left(1-\frac{m_\yvec}{c}\right) P(m_\xvec+1,m_\yvec;t)	\nonumber\\
+\alpha \, m_\yvec\left(1-\frac{m_\xvec-1}{c}\right) P(m_\xvec-1,m_\yvec;t)	\nonumber\\
 -\alpha \left[ m_\xvec\left(1-\frac{m_\yvec}{c}\right) + m_\yvec \left(1-\frac{m_\xvec}{c}\right)  \right]P(m_\xvec,m_\yvec;t)\Big\}\, , 
\elabel{mastEq1}
\end{eqnarray}
where $P(\{m\};t)$ is the probability of finding the system in microstate $\{m\}$ at time $t$,
$\xvec$ and $\yvec$ are nearest neighbours, and $c$ is the local carrying capacity, the maximum number of particles
allowed in each site. This carrying capacity implements the fermionic nature of opinions and may be set to $c=1$.
However, it is useful to parametrise this excluded volume constraint as it is the only interaction that keeps the spreading 
of particles at bay \cite{NekovarPruessner:2016,bordeu2019volume,vanWijland:2001}.
The first term in \Eqref{mastEq1} describes the elimination of a particle at $\xvec$ by an empty space at $\yvec$; and
the second, the production of a particle at $\xvec$ by a particle at $\yvec$.
Initially, at time $t_0=0$, a single particle is placed at $\xvec_0$.
Following the Doi-Peliti formalism, the master equation in \eref{eq:mastEq1} is cast in path integral form \cite{Taeuber:2014}.
After taking the continuum limit and introducing the annihilation field $\phi(\xvec,t)$ and the Doi-shifted creation field 
$\tildephi(\xvec,t)$, the action functional
of the resulting field theory is $\mA_0+\mA_\textnormal{\scriptsize int}$, where 
\numparts
\begin{eqnarray}
\elabel{VMmA}
\elabel{VMmA0}
\fl\mA_0 &= \int \plaind^d\xvec\plaind t \Big\{ 
\tildephi \partial_t\phi -D \tildephi\nabla^2\phi + r\tildephi \phi
\Big\} \, ,\\
\elabel{VMmA1}
\fl\mA_\textnormal{\scriptsize int} &= \int \plaind^d\xvec\plaind t \Big\{ 
- {s}\tildephi^2\phi + {\chi} \tildephi^2\phi^2 
 + w_1 \tildephi \left(\nabla^2{\phi}\right) \tildephi
 - w_2 \tildephi \left(\nabla^2{\phi}\right) \tildephi^2 \phi + w_3 \tildephi^3\phi^2
\Big\} \,,
\end{eqnarray}
\endnumparts
$D=\alpha$ is the effective diffusion constant,
${s} =\alpha d$ is the effective branching rate,
and the other non-linearities are
${\chi}=\alpha d/c$,
$w_1 = \alpha$,
$w_2 =\alpha/c$ and
$w_3 = \alpha d$.
The mass $r>0$ of the particles  
is added to regularise the infrared, ensuring causality in the time domain $t\geq t_0$,
and is later removed by taking the limit $r\to0$ at criticality.
In fact, if the two processes $A\to A+A$ and $A\to\emptyset$ in Fig.~\ref{VM04} are set to happen at rates 
$\alpha_1$ and $\alpha_2$ respectively, with $\alpha_1\neq\alpha_2$, a mass term $(\alpha_2-\alpha_1)d\tildephi\phi$
emerges  in the action naturally.
The action functional allows the calculation of an observable $\mO$,
via the path integral
\begin{equation}
\elabel{expectation}
\ave{\mO} = \int \mD [\phi,\tildephi] \, \mO \, \exp{-(\mA_0+\mA_\textnormal{\scriptsize int})} \, ,
\end{equation}
which satisfies the normalisation condition $\ave{1}=1$.

The long-range behaviour in space and time of the spreading of particles in the VM is governed by
the  processes of effective branching and effective diffusion. 
Consequently, I demand that $s$ and $D$ have independent dimensions, 
 $\left[{s}\right]=\Adim$ and $[D] = \Bdim$. This choice of engineering dimensions is ultimately a 
decision about how the continuum limit is taken, which, in the present case, is done by
demanding that neither branching nor diffusion are irrelevant in any dimension.
Along with $[\xvec]=\Ldim$ and the dimensionlessness of the action functional, 
$[\mA_0+\mA_\textnormal{\scriptsize int}]=1$, it follows that
\begin{eqnarray}
\left[  \phi \right] =  \Ldim^{-d+2}\Adim\Bdim^{-1}, \quad
[  \tildephi ] =  \Ldim^{-2}\Adim^{-1}\Bdim , \quad
\left[  t \right] = \Ldim^{2}\Bdim^{-1},  \elabel{dimensionsVM}\\ \fl\quad
\left[  r \right] = \Ldim^{-2}\Bdim, \quad
\left[  {\chi} \right] = \Ldim^{d-2}\Bdim, \quad
\left[  w_1 \right] = \Ldim^2\Adim, \quad
\left[  w_2 \right] = \Ldim^{d+2} \Adim, \quad
\left[  w_3 \right] = \Ldim^d\Adim , \nonumber
\end{eqnarray}
implying that the couplings $w_1$, $w_2$ and $w_3$ are irrelevant, which yields the action functional in \Esref{VMmA}
and \eref{eq:VMmA1}
 consistent with \cite{Dickman:1995,Munoz:1997}. 
  Moreover, the action of the VM is equivalent to the action of the diffusion-limited pair annihilation process \cite{peliti1986renormalisation} under rapidity reversal \cite{Taeuber:2014},
  that is under time inversion and the exchange of fields
 \begin{equation}
 \phi (x,t) \leftrightarrow -\tildephi(x,-t) \,.
 \elabel{rap_rev}
 \end{equation}
As a result, the renormalisations of both field theories are intimately related. 
  The choice of engineering dimensions above is not unique.
 An alternative  choice \cite{peliti1986renormalisation} is, for example, $[s]=[\chi]$.

 The coupling ${\chi}$ has critical 
dimension $d_c=2$ such that ${\chi}$ is relevant for $d<d_c$, marginal at $d=d_c$ and irrelevant for $d>d_c$.
The fact that, in terms of the original parameters, the coupling $\chi$ is inversely proportional to the carrying capacity $c$
indicates that $\chi$ arises from the excluded-volume interactions between particles, which, according to the critical 
dimension $d_c$ become irrelevant for dimensions $d>d_c$.

The Fourier transform is defined, by convention, as
\numparts
\begin{eqnarray}
\phi(\kvec,\omega) &=& \int \plaind^d\xvec \plaind\plaint \, \exp{\imag\omega t-\imag\kvec \xvec}\phi(\xvec,t) \, ,\\
\phi(\xvec,t) &=& \int \dbar^d\kvec\dbar\omega \, \exp{-\imag\omega t+\imag\kvec\xvec}\phi(\kvec,\omega) \, ,
\end{eqnarray}
\endnumparts
where the spatial momentum $\kvec$ is  conjugate of the position $\xvec$,
 the frequency $\omega$ is  conjugate of time $t$,
 $\dbar^d\kvec = \plaind^d\kvec /(2\pi)^d $ and $\dbar\omega = \plaind\omega/2\pi$.
 The same convention applies to $\tildephi(\xvec,t)$.
It is often convenient to use Feynman diagrams to express lengthy expressions \cite{Taeuber:2014}.
In the diagrams, time, or the direction of causality {as implemented by the positive mass $r$},
is read from right to left. The bare propagator of the $\phi$ field is  then, from \Eqref{VMmA0},
\begin{eqnarray}
\tikz[baseline=-2.5pt]{
\node at (0.5,0) [above] {$\tildephi$};
\node at (-0.5,0) [above] {$\phi$};
\draw[Aactivity] (0.5,0) -- (-0.5,0) node[at end,above] {};
}\,
&\hat{=}& \ave{\phi(\kvec,\omega)\tildephi(\kvec',\omega')} \\
&=& \frac{\deltabar(\omega+\omega')\deltabar(\kvec+\kvec')}{-\imag \omega  + D \kvec^2 + r} \, .\nonumber
\end{eqnarray}

\subsection{Exponents $\eta$ and $z$}
The expected number of particles in the system is
\begin{equation}
n(t) = \int \plaind^d\xvec \ave{\phi(\xvec,t)\tildephi(\xvec_0,t_0)} 
=  \Theta(t-t_0)\exp{-r (t-t_0)} \, ,
\end{equation}
where $\Theta$ is the Heaviside step function and $r>0$. At criticality, $r=0$, the expected number of particles is $n(t)=1$, 
so that $\eta=0$ according to \Eqref{crit_exp}.
The mean square spread of particles is
\begin{eqnarray}
\elabel{MSD1}
\fl \quad
R^2(t) = \int \plaind^d\xvec (\xvec-\xvec_0)^2 \ave{\phi(\xvec,t)\tildephi(\xvec_0,t_0)} \\
\fl\quad\quad\quad
= -   \frac{\plaind ^2}{\plaind \kvec^2} \, \left( \int \dbar\kvec'\dbar\omega\dbar\omega' \, 
\exp{-\imag\omega t+\imag\kvec\xvec} \exp{-\imag\omega' t_0+\imag\kvec'\xvec_0} 
\ave{\phi(\kvec,\omega)\tildephi(\kvec',\omega')}\right)\Big|_{\kvec=\gpvec{0}} \nonumber\\
\fl\quad\quad\quad
= {2dD} t \exp{- r t}  \,, \nonumber
\end{eqnarray}
where $t_0=0$ and
 $r>0$. At the critical point, $R^2(t)\propto t$, so that $z=1$. 
The reason why both exponents, $\eta$ and $z$, attain their trivial values is the non-renormalisation
of the propagator and the absence of a temporal singularity at time $t=t_0$ that would require an independent
renormalisation constant \cite{Taeuber:2014,JanssenTaeuber:2005}.

\subsection{Exponent $\delta$}\label{epxdeltaVM}
According to  \cite{JanssenTaeuber:2005, Janssen:2005, Garcia2018branchings}, the survival probability is given by
\numparts
\begin{eqnarray}
\Ps (r,D,s,\chi;t) = -\ave{\exp{-\int\plaind^d\xvec \phi(\xvec,t)}\tildephi(\xvec_0,t_0)} \\
\quad= -\sum_{n=0}^\infty \frac{(-1)^{n}}{n!} \ave{\left({\int\plaind^d\xvec \phi(\xvec,t)}\right)^n\tildephi(\xvec_0,t_0)}\\
\quad\hat{=} -\sum_{n=0}^\infty \frac{(-1)^{n}}{n!}\quad
\tikz[baseline=-2.5pt]{
\begin{scope}[rotate=-30]
  \draw [decorate,decoration={brace,amplitude=5pt}] (-128:1.4cm) -- (-172:1.4) node[pos=0.5,left,xshift=-0.07cm,yshift=0.14cm] {$n$};
  \draw[Aactivity] (-130:0.3) -- (-130:1.3);
  \path [postaction={decorate,decoration={raise=0ex,text along path,
  text={|\large|....}}}]
  (-155:1.2cm) arc (-155:-130:1.2cm);
  \draw[Aactivity] (-155:0.3) -- (-155:1.3);
  \draw[Aactivity] (-170:0.3) -- (-170:1.3);
\end{scope}
\draw[thick,fill=white,pattern=north east lines] (0,0) circle (0.3cm);
  \draw[Aactivity] (0:0.3) -- (0:0.8);
} \elabel{VM_treediagram}
\end{eqnarray}
\endnumparts
 where the hatched circle in \Eqref{VM_treediagram} indicates the sum over all  diagrams that have 
 one incoming leg
 and $n$ outgoing legs. For instance, the two tree-level contributions to $n=4$ are
 \begin{equation}
\tikz[baseline=-2.5pt]{
\draw[Aactivity] (0.5,0) -- (0,0);
\draw[Aactivity] (140:0.8) -- (0,0);
\draw[Aactivity] (-140:0.8) -- (0,0);
\draw[Aactivity] (-140:0.4)+(160:0.35) -- (-140:0.4);
\draw[Aactivity] (140:0.4)+(-160:0.35) -- (140:0.4);}
\, , \quad
\tikz[baseline=-2.5pt]{
\draw[Aactivity] (0.5,0) -- (0,0);
\draw[Aactivity] (140:0.8) -- (0,0);
\draw[Aactivity] (-140:0.8) -- (0,0);
\draw[Aactivity] (140:0.5)+(-140:0.3) -- (140:0.5);
\draw[Aactivity] (140:0.25)+(-140:0.55) -- (140:0.25);}
\, .
\end{equation}
At tree level, the sum \eref{eq:VM_treediagram} is easily performed to produce
\begin{equation}
\Ps (r,D,s,\chi;t) = \frac{\exp{-rt}}{1+\frac{s}{r}\left(1-\exp{-rt}\right)} \, ,
 \elabel{psurvVM}
\end{equation}
which becomes
\begin{equation}
\Ps (0,D,s,\chi;t) = \frac{1}{1+st}
\elabel{MFTcrit}
\end{equation}
in the limit of $r\to0$. However, in $d<2$, where $\chi$ is relevant, loop diagrams contribute to the survival probability,
whose effect is captured by replacing the bare $s$ in \eref{eq:psurvVM} by an effective (renormalised) coupling.
 This non-linearity renormalises under dimensional regularisation at $d=2-\epsilon$, by means
of the governing non-linearity ${\chi}$, which renormalises itself.

The effective, renormalised, couplings 
$\chi_{\textnormal{\scriptsize eff}}$ and $s_{\textnormal{\scriptsize eff}}$ are, diagrammatically,
\numparts
\begin{eqnarray}
\elabel{renormVM_diag}
\tikz[baseline=-2.5pt]{
\draw[Aactivity] (140:0.5) -- (-40:0.5);
\draw[Aactivity] (-140:0.5) -- (40:0.5);
\draw[black,fill=black] (0,0) circle (0.1) node[at end,above=3pt] {$\chi_{\textnormal{\scriptsize eff}}$};
}
=
\tikz[baseline=-2.5pt]{
\node [above=3pt] {${\chi}$};
\draw[Aactivity] (140:0.5) -- (-40:0.5);
\draw[Aactivity] (-140:0.5) -- (40:0.5);
}
+
\tikz[baseline=-2.5pt]{
\draw[Aactivity] (0,0) circle (0.4);
\node [above=3pt,xshift=14pt] {${\chi}$};
\node (-0.5,0) [above=3pt,xshift=-14pt] {${\chi}$};
\draw[Aactivity] ($(140:0.4)+(-0.4,0)$) -- (-0.4,0);
\draw[Aactivity] ($(-140:0.4)+(-0.4,0)$) -- (-0.4,0);
\draw[Aactivity] ($(40:0.4)+(0.4,0)$) -- (0.4,0);
\draw[Aactivity] ($(-40:0.4)+(0.4,0)$) -- (0.4,0);
} 
+
\tikz[baseline=-2.5pt]{
\draw[Aactivity] ($(140:0.4)+(-0.4,0)$) -- (-0.4,0);
\draw[Aactivity] ($(-140:0.4)+(-0.4,0)$) -- (-0.4,0);
\node (-0.5,0) [above=3pt,xshift=-14pt] {${\chi}$};
\draw[Aactivity] (0,0) circle (0.4);
\node (0.4,0) [above=8pt,xshift=11pt] {${\chi}$};
\draw[Aactivity] (0.8,0) circle (0.4);
\draw[Aactivity] ($(40:0.4)+(1.2,0)$) -- (1.2,0) node [above=3pt,xshift=2pt] {${\chi}$};
\draw[Aactivity] ($(-40:0.4)+(1.2,0)$) -- (1.2,0);
} \,
+\ldots \elabel{chi_renorm}\\
\tikz[baseline=-2.5pt]{
\draw[Aactivity] (0.5,0) -- (0,0) node[at end,above=3pt,xshift=3pt] {$s_{\textnormal{\scriptsize eff}}$};
\draw[Aactivity] (130:0.5) -- (0,0);
\draw[Aactivity] (-130:0.5) -- (0,0);
\draw[black,fill=black] (0,0) circle (0.1);
}
=
\tikz[baseline=-2.5pt]{
\draw[Aactivity] (0.5,0) -- (0,0) node[at end,above=3pt,xshift=3pt] {${s}$};
\draw[Aactivity] (130:0.5) -- (0,0);
\draw[Aactivity] (-130:0.5) -- (0,0);
}
+
\tikz[baseline=-2.5pt]{
\draw[Aactivity] (0,0) circle (0.4);
\draw[Aactivity] (0.8,0) -- (0.4,0) node[at end,above=3pt,xshift=4pt] {${s}$};
\node (-0.5,0) [above=3pt,xshift=-14pt] {${\chi}$};
\draw[Aactivity] ($(140:0.4)+(-0.4,0)$) -- (-0.4,0);
\draw[Aactivity] ($(-140:0.4)+(-0.4,0)$) -- (-0.4,0);
} \,
+
\tikz[baseline=-2.5pt]{
\draw[Aactivity] ($(140:0.4)+(-0.4,0)$) -- (-0.4,0);
\draw[Aactivity] ($(-140:0.4)+(-0.4,0)$) -- (-0.4,0);
\node (-0.5,0) [above=3pt,xshift=-14pt] {${\chi}$};
\draw[Aactivity] (0,0) circle (0.4);
\node (0.4,0) [above=8pt,xshift=11pt] {${\chi}$};
\draw[Aactivity] (0.8,0) circle (0.4);
\draw[Aactivity] (1.2,0) node[above=3pt,xshift=4pt] {${s}$} -- (1.6,0) ;
} \,
+\ldots  \elabel{s_renorm}
\end{eqnarray}
\endnumparts
which can be summed exactly,
\numparts
\begin{eqnarray}
\elabel{renormVM}
\chi_{\textnormal{\scriptsize eff}}= \frac{\chi}{1+\mJ \chi}\,,\\
s_{\textnormal{\scriptsize eff}} = \frac{s}{1+\mJ \chi} \,,
\end{eqnarray}
\endnumparts
taking into account that $s$ enters in the expansion of the exponential of the action in \Eqref{expectation} with a positive sign
and $\chi$ with a negative sign, and
where $\mJ$ is the loop integral 
\numparts
\begin{eqnarray}
\elabel{loopVM}
\fl\mJ =&
\tikz[baseline=-2.5pt]{
\draw[Aactivity] (0,0) circle (0.4);
\draw[Aactivity] (0.5,0) -- (0.4,0);
\draw[Aactivity] (-0.5,0) -- (-0.4,0);
}\\
\fl \corresponding& \int\dbar^d\kvec_1\dbar^d\kvec_2\dbar^d\kvec_3\dbar^d\kvec_4
 \dbar\omega_1\dbar\omega_2\dbar\omega_3\dbar\omega_4
 \, \deltabar(\kvec_2+\kvec_4)
\deltabar(\kvec_1+\kvec_3)
 \nonumber\\ \fl& \times
\deltabar(\omega_2+\omega_4)
\deltabar(\omega_1+\omega_3)
\ave{\phi(\kvec_2,\omega_2)\tildephi(\kvec_1,\omega_1)}
\ave{\phi(\kvec_4,\omega_4)\tildephi(\kvec_3,\omega_3)} \nonumber\\
\fl=&
\frac{r^{\frac{d}{2}-1}}{(4\pi D)^{\frac{d}{2}}}  \Gamma\left(1-\frac{d}{2}\right)
\simeq \frac{2 r^{-\frac{\epsilon}{2}}}{\epsilon(4\pi D)^{\frac{d}{2}}} \, , 
\end{eqnarray}
\endnumparts
 with $d=2-\epsilon$. 

 Defining the $Z$-factors such that $\chi_{\textnormal{\scriptsize eff}}=\chi Z_\chi$ and $s_{\textnormal{\scriptsize eff}}=s Z_s$, from \Eqref{renormVM} it follows 
 that the two $Z$-factors are equal, $ Z_\chi=Z_s$, 
 yielding the renormalisation of the coupling $s$ entirely driven by the renormalisation of $\chi$.
It is noteworthy that the renormalisation of this pair of couplings satisfies the Ward identity \cite{bordeu2019volume}
 \begin{equation}
 \elabel{W_ID}
 \frac{\plaind s_{\textnormal{\scriptsize eff}}}{\plaind s} =  \frac{\chi_{\textnormal{\scriptsize eff}}}{\chi}  \, ,
 \end{equation}
 derived in  \ref{Ward}.
 
 The renormalisation of the dimensionless coupling $\chi_\renorm=\mu^{-\epsilon} \mU {\chi_{\textnormal{\scriptsize eff}}}$,
  where $\mu=\sqrt{r/D}$ is an arbitrary inverse length scale and $\mU=2/\left(D(4\pi)^{{d}/{2}}\right)$,
  is, using \Erefs{renormVM}, \eref{eq:loopVM} and the identity $x=y/(1+ay)=y(1-ax)$, exactly
\begin{equation}
\chi_\renorm = \mu^{-\epsilon} \mU {\chi} \left( 1-\frac{1}{\epsilon} \chi_\renorm \right)  \, .
\end{equation}
The $\beta$-function of the renormalised governing coupling $\chi_\renorm$ is
\begin{equation}
\beta_{\chi} = \mu\frac{\plaind}{\plaind\mu} \chi_\renorm = -\epsilon\chi_\renorm + \chi_\renorm^2 \, ,
\end{equation}
whose infrared-stable fixed point is $\chi_\renorm^*=\epsilon$ since 
$ \plaind\beta_\chi(\chi_\renorm^*=\epsilon)/\plaind\chi_\renorm=\epsilon>0$.
Since both Z-factors are identical, $Z_\chi=Z_{s}=1- {\chi_\renorm}/{\epsilon}$,
the Wilson $\gamma$-functions of the couplings are then identical as well,
\begin{equation}
\gamma_{s} = \gamma_\chi=\mu\frac{\plaind}{\plaind\mu}\log Z_\chi
=\frac{-1}{\epsilon Z_\chi}\beta_\chi =\chi_\renorm \to \chi_\renorm^* \, .
\elabel{WilsonVM}
\end{equation}

Then, the anomalous dimensions of $s$ and $\chi$ are $\gamma_s=\gamma_\chi=\epsilon$ for $\epsilon>0$.
For $\epsilon<0$, the infrared-stable fixed point of $\beta_\chi$ is $\chi^*=0$ as 
$  \plaind\beta_\chi(\chi_\renorm^*=0)/\plaind\chi_\renorm=-\epsilon>0$,
and therefore $\gamma_s=\gamma_\chi=0$. As a result, in dimensions $d>2$ the anomalous dimensions of $s$ and $\chi$
 vanishes and the process exhibits mean-field behaviour as correlations due to volume exclusion become irrelevant. 
 
 Dimensional consistency in \Eqref{psurvVM} requires $[s]=[r]$ and therefore $\Adim=\Ldim^{-2}\Bdim$.
 As $[\Psurv]=1$ it follows that
 \begin{equation}
\elabel{dim_ana}
\Ps (r,D,s,\chi;t)  = \Psurv\left(\frac{r}{L^{-2}B}, \frac{D}{B}, \frac{s}{L^{-2}B}, \frac{\chi}{L^{-\epsilon}B}; \frac{t}{L^{2}B^{-1}} \right) \, , 
 \end{equation}
 for any positive, real $L$, $B$. 
 A similar expression can be found for all vertex functions. Following the usual process \cite{Taeuber:2014} of using
 \eref{eq:dim_ana} with $B=D$ and $L=\sqrt{Dt}$ in conjunction with the solution of the Callan Symanzik equation
 on the basis of the Wilson $\gamma$-functions \eref{eq:WilsonVM} produces
 \begin{equation}
 \Ps (r,D,s,\chi;t) = \Psurv\left(rt, 1, st \ell^{\gamma_s}, \frac{\chi\ell^{\gamma_\chi}}{D(Dt)^{-\epsilon/2}}; 1 \right) \, ,
 \end{equation}
 with a dimensionless, small scale parameter $\ell$, which one may choose to be 
 $\ell=\left( D/\chi \cdot(Dt)^{-\epsilon/2}\right)^{1/\gamma_\chi}$, such that at $\epsilon>0$,
 when $\gamma_s=\gamma_\chi=\epsilon$,
  \begin{equation}
 \Ps (r,D,s,\chi;t) = \Psurv\left(rt, 1, \frac{s}{\chi}(Dt)^{1-\frac{\epsilon}{2}}, 1; 1 \right) \, .
 \end{equation}
 As the accounts for the effect of the interaction that takes hold at large times by adjusting parameters in an
 expression involving (arbitrarily) short times, when the tree-level theory applies, the survival probability at criticality
\eref{eq:MFTcrit} is expected to be 
 \begin{equation}
 \Ps (0,D,s,\chi;t) = \frac{1}{1+\frac{s}{\chi}(Dt)^{1-\frac{\epsilon}{2}}} \, .
 \end{equation}
Asymptotically in $t$,
\begin{equation}
\Ps (0,D,s,\chi;t) \propto t^{-\left(1-\frac{\epsilon}{2}\right)} \, ,
\end{equation}
which implies by \Eqref{crit_exp} that $\delta = 1-{\epsilon}/{2}$ for $d<2$. 
For $d\geq2$ the $\gamma$-functions vanish so that
$\delta=1$, with logarithmic corrections at $d=d_c=2$, recovering the results in \cite{Dickman:1995}.

\section{Critical exponents of the Concealed Voter Model \label{FT_CVM}}
\begin{figure}
 \centering
\subfigure[]{ \label{CVM01} \includegraphics[]{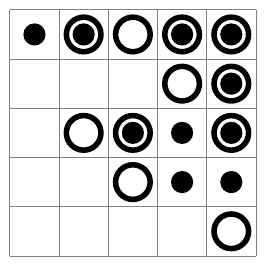}}
\subfigure[]{ \label{CVM06} \includegraphics[]{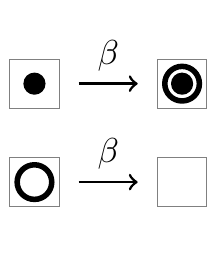}}
\subfigure[]{ \label{CVM07} \includegraphics[]{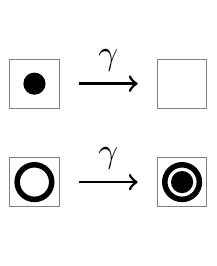}}
\captionsetup{justification=raggedright,singlelinecheck=false}
\caption{\label{CVM} The concealed voter model
(opinions in the outer layer are represented by circles and empty spaces, and,
 in the inner layer, by solid disks and empty spaces).
\subref{CVM01} Sample of a particle configuration in the CVM in $d=2$.
\subref{CVM06} Externalisation process between the two layers.
\subref{CVM07} Internalisation process between the two layers.
The process of copying in the outer layer is represented in Fig.~\ref{VM04}.}
\end{figure}

In addition to the lattice (outer layer) of the voter model above, where opinions are copied to nearest neighbours
with rate $\alpha$, the CVM has an inner layer, that represents the private opinion of each agent, 
which may equal or differ from the expressed opinion in the outer layer, Fig.~\ref{CVM}. 
The private and publicly expressed opinions interact via two new processes:
an agent may \textbf{externalise} their private opinion with rate $\beta$
 (the opinion in the inner layer is copied into the outer layer)
and \textbf{internalise} their expressed opinion with rate $\gamma$
(the opinion in the outer layer is copied into the inner layer). In total, there are three concurrent Poisson
processes: copying at
rate $\alpha$, externalisation at rate $\beta$ and internalisation at rate $\gamma$. 
The original VM is recovered if $\beta=0$ \cite{Gastner_2018}.

As in the VM, the interaction between two competing opinions can be represented by the interactions between particles
and empty spaces. The particles in the two layers may be considered two distinct species, namely,
mobile particles with occupation numbers $\{m\}$ 
and immobile particles with occupation numbers $\{n\}$ \cite{NekovarPruessner:2016, bordeu2019volume}.

The master equation of the CVM comprises in addition to \Eqref{mastEq1} 
 for the process of copying (the left-hand side of the equation becomes
$\dot{P}_{\alpha} (\{m\};t)$)
the following contribution 
\begin{eqnarray}\elabel{mastEq2}
\fl\dot{P}_{\beta} (\{m\},\{n\};t) = 
\sum_\xvec &\Big\{  \beta (m_\xvec+1)\left(1-\frac{n_\xvec}{c}\right) P(n_\xvec,m_\xvec+1;t) \nonumber\\
&+ \beta n_\xvec\left(1-\frac{m_\xvec-1}{c}\right) P(n_\xvec,m_\xvec-1;t) \nonumber\\
&-\beta \left[ m_\xvec\left(1-\frac{n_\xvec}{c}\right) + n_\xvec\left(1-\frac{m_\xvec}{c}\right) \right]
 P(n_\xvec,m_\xvec;t)\Big\},
\end{eqnarray}
 for the process of externalisation with rate $\beta$, 
and a symmetric transformation of the above for the process of internalisation obtained by 
swapping $m_\xvec$ for $n_\xvec$, the rate $\beta$ for $\gamma$. 
Hence, the time evolution of the microstate probability is
$\dot{P}(\{m\},\{n\};t) = \dot{P}_{\alpha} (\{m\};t)
+\dot{P}_{\beta} (\{m\},\{n\};t)+\dot{P}_{\gamma} (\{m\},\{n\};t)$.
At time $t_0=0$, the system is initialised with one particle of either species with equal probability at $\xvec_0$.

It follows that the action functional $\mA_0'+\mA_\textnormal{\scriptsize int}'$ of the CVM is
\numparts
\begin{eqnarray}
\fl \mA_0' = \int \plaind^d\xvec\plaind t \Big\{ 
\tildephi \partial_t\phi -D \tildephi\nabla^2\phi + r_1\tildephi \phi
+\tildepsi \partial_t\psi + r_2\tildepsi \psi
-\tau_1 \psi\tildephi
-\tau_2 \phi\tildepsi
\Big\}\elabel{CVM_A0}\\
\fl \mA_\textnormal{\scriptsize int}' = \int \plaind^d\xvec\plaind t \Big\{ 
- {s}\tildephi^2\phi 
-\sigma_1\tildepsi\psi\tildephi -\sigma_2\tildephi\phi\tildepsi 
+ {\chi} \tildephi^2\phi^2 
+\kappa_1\psi\tildephi^2\phi +\kappa_2\phi\tildepsi^2\psi\\
 + w_1 \tildephi \left(\nabla^2{\phi}\right) \tildephi\nonumber
 - w_2 \tildephi \left(\nabla^2{\phi}\right) \tildephi^2 \phi + w_3 \tildephi^3\phi^2
 +w_4\tildepsi\psi\tildephi^2\phi + w_5\tildephi\phi\tildepsi^2\psi\Big\}
\nonumber
\end{eqnarray}
\endnumparts
where $\psi$ is the annihilation and $\tildepsi$ is the Doi-shifted creation field of the immobile species, and where,
in terms of the model parameters, $r_1=\tau_1=\sigma_1=\beta$ (associated with externalisation),
 $r_2=\tau_2=\sigma_2=\gamma$ (associated with internalisation), 
$\kappa_1=w_4=\beta/c$ and $\kappa_2=w_5=\gamma/c$. 
The mass terms $r_1$ and $r_2$ emerge from  the microscopic dynamics of the process:
$r_1$ derives from the depletion of mobile particles due to externalisation (Fig.~\ref{CVM06} bottom) and
$r_2$  from the depletion of immobile particles due to internalisation (Fig.~\ref{CVM07} up).
The couplings $\tau_1$ and $\tau_2$ are commonly referred to as transmutations
\cite{NekovarPruessner:2016,bordeu2019volume},
as they provide causal correlations between fields of different species (Fig.~\ref{CVM06} up and Fig.~\ref{CVM07} bottom,
respectively). This terminology may lead to confusion, though, if one thinks of particles in one species changing 
into the other species. In the CVM it is more accurate to say that 
 immobile particles deposit mobile particles with rate $\beta$
 and mobile particles deposit immobile particles with rate $\gamma$.
Assuming that $\Ldim$, $\Adim$ and $\Bdim$
are independent dimensions, and that the two fields $\phi$ and $\psi$ have the same dimensions,
$\left[  \phi \right] = \left[  \psi \right]  = \Ldim^{-d+2}\Adim\Bdim^{-1}$, the dimensions
of the new couplings are
\begin{eqnarray}
 \left[  r_1 \right] = \left[  r_2 \right] = \left[  \tau_1 \right] = \left[ \tau_2 \right] = \Bdim\Ldim^{-2} \, , \quad
\left[  \sigma_1 \right] = \left[  \sigma_2 \right] = \Adim \, , \\
\left[  \kappa_1 \right] =\left[  \kappa_2 \right]= \Ldim^{d-2}\Bdim \, , \quad
\left[  w_4 \right] = \left[  w_5 \right]= \Ldim^d\Adim \, . \nonumber
\end{eqnarray}
This implies that $w_4$ and $w_5$ are irrelevant non-linearities, while all the others are either relevant or marginal.

The couplings $s$ and $\chi$ stay as in the VM, 
but  the new couplings 
$\sigma_1$, $\sigma_2$, $\kappa_1$ and $\kappa_2$ couple the new ${\psi}$, ${\tildepsi}$ fields to the old  ${\phi}$, 
${\tildephi}$.
Diagrammatically,
\begin{equation}
\tikz[baseline=-2.5pt]{
\draw[substrate] (0.5,0) -- (0,0) node[at end,above=3pt,xshift=3pt] {${\sigma_1}$};
\draw[substrate] (130:0.5) -- (0,0);
\draw[Aactivity] (-130:0.5) -- (0,0);
},\quad
\tikz[baseline=-2.5pt]{
\draw[Aactivity] (0.5,0) -- (0,0) node[at end,above=3pt,xshift=3pt] {${\sigma_2}$};
\draw[Aactivity] (130:0.5) -- (0,0);
\draw[substrate] (-130:0.5) -- (0,0);
},\quad
\tikz[baseline=-2.5pt]{
\node [above=3pt] {${\kappa_1}$};
\draw[Aactivity] (140:0.5) -- (0,0);
\draw[substrate] (0,0) -- (-40:0.5);
\draw[Aactivity] (-140:0.5) -- (40:0.5);
},\quad
\tikz[baseline=-2.5pt]{
\node [above=3pt] {${\kappa_2}$};
\draw[substrate] (140:0.5) -- (0,0);
\draw[Aactivity] (0,0) -- (-40:0.5);
\draw[substrate] (-140:0.5) -- (40:0.5);
},
\elabel{couplCVM}
\end{equation}
where \tikz[baseline=-2.5pt]{ \draw[substrate] (0.8,0) -- (0,0) ;} represents the propagator $\ave{\psi\tildepsi}$.
This set of mixed couplings encode correlations between the two species due to transmutation and volume exclusion.
From \Eqref{CVM_A0}, the four response propagators $\ave{\phi\tildephi}$, $\ave{\psi\tildepsi}$,
$\ave{\phi\tildepsi}$ and $\ave{\psi\tildephi}$ are determined,  \ref{app_propsCVM1}.

In fact, all six $s$-like couplings (one incoming leg and two outgoing legs) are effectively available by combining $s$,
$\sigma_1$ or $\sigma_2$ with suitable propagators, and are denoted by $\sigma_3$, \ldots, $\sigma_5$.
Similarly, the nine $\chi$-like couplings (two incoming and two outgoing legs) are available, by combining
$\chi$, $\kappa_1$ or $\kappa_2$ with the propagators, and are denoted by $\kappa_3$, \ldots, $\kappa_8$.
Therefore, the one-loop renormalisation of each of the available couplings may need to consider sixteen loop integrals, 
which are reduced to ten integrals due to their symmetries,  \ref{app_loop_integrals}.

However, to define the observables $n(t)$, $R^2(t)$ and $\Psurv(t)$ in the CVM, it is convenient
 to consider the total number of particles, irrespectively of their species.
 Indeed, the system will reach an absorbing state whenever the total population is zero.
This motivates the definition of the annihilation \emph{density} field $\rho=\phi+\psi$
 (and the Doi-shifted creation density field 
$\tilderho=\tildephi+\tildepsi$), whose propagator is, using the propagators in  \ref{app_propsCVM1},
\begin{eqnarray}
\tikz[baseline=-2.5pt]{
\node at (0.5,0) [above] {$\tilderho$};
\node at (-0.5,0) [above] {$\rho$};
\draw[density] (0.5,0)  -- (-0.5,0) ;
} & \hat{=} \ave{\rho(\kvec,\omega)\tilderho(\kvec',\omega')}\\
& = 
\deltabar(\omega+\omega')\deltabar(\kvec+\kvec') 
\frac{
-2\imag \omega + D \kvec^2 + r_1 +r_2 +\tau_1 +\tau_2}{\left(-\imag \omega  + D \kvec^2 + r_1\right)\left(-\imag \omega  +r_2\right)-\tau_1\tau_2}  \,. \nonumber
\end{eqnarray}

The critical point of the CVM happens when particles, \ie one of the two opinions, are no longer subject to
\emph{spontaneous} extinction, $r_1=\tau_1$ and $r_2=\tau_2$.
That is, when the gain and loss of particles due to  deposition and decay are balanced:
immobile particles through the internalisation process, and mobile particles
through the externalisation process.
To deal with the infrared divergence at criticality,
 it can be assumed that $r_1=\tau_1+{r}$ and $r_2=\tau_2+{r}$, taking the limit ${r}\to0$ to reach the critical point.
   This parametrisation of $r_1$ and $r_2$ is allowed because the critical point  $r\to0$ of the CVM is unique,
 as shown in \ref{app_criticalpoint}.

 The loop integrals in  \ref{app_loop_integrals} contain two infrared divergencies,
 $r\to0$ and $\tau_1+\tau_2\to0$, where the second one corresponds to the original VM at criticality.
 Henceforth, I assume $\tau_1+\tau_2>0$ and refer to $r\to0$ as the critical point of the CVM.
 Moreover, the loop integral $I_{10}$ in \Eqref{CVM_I3} has a quadratic divergence that results from taking the 
 continuum limit. Since the process is originally defined on the lattice, this divergence can be regularised by
 introducing an ultraviolet cutoff. Usually, such a quadratic divergence can be absorbed into a fluctuation-induced shift of 
 the critical point by means of additive renormalisation. However, in this case the quadratic divergence is
 associated with an infrared divergence in $\tau_2$, \ie the critical point of the original VM within the CVM.

\subsection{Exponents $\eta$ and $z$}
The expected number of particles is
\begin{equation}\elabel{CVM_nt}
 n(t)=\frac{1}{2} \int \plaind^d\xvec \ave{ \rho(\xvec,t)  \tilderho(\xvec_0,t_0)}  = \Theta(t-t_0)\exp{-r (t-t_0)} \,,
 \end{equation}
 where the factor of $1/2$ accounts for a particle of either species being created with equal probability.
 At $r=0$, \ie at the critical point $n(t)=1$, and the critical exponent in \Eqref{crit_exp} is $\eta=0$.
The mean square spread of mobile particles is
\begin{eqnarray}
\elabel{CVM_R2}
R^2(t) & = \frac{1}{2}\int \plaind^d\xvec (\xvec-\xvec_0)^2 \ave{\rho(\xvec,t)\tilderho(\xvec_0,t_0)} \\
& = \frac{dD}{\tau_1+\tau_2} \exp{-{r} t} \Big[2\tau_2 t  + \frac{\tau_1-\tau_2}{\tau_1+\tau_2} 
\left(1-\exp{-(\tau_1+\tau_2)t}\right)\Big] \, , \nonumber
\end{eqnarray}
where $t_0=0$.
At the critical point $R^2(t)\propto t$, so the critical exponent is $z=1$.

\Eqref{CVM_R2} recovers the behaviour of the original voter model, \Eqref{MSD1}, for vanishing transmutation, depending
on how this limit is taken.
For $\tau_1=0$, $\tau_2\to0$ and $\tau_2=0$, $\tau_1\to0$ the mean square spread is half of that of \Eqref{MSD1},
as the spread is zero if the initial particle happens to be immobile.

\subsection{Exponent $\delta$}
The survival probability in the CVM is
\numparts
\begin{eqnarray}
\elabel{psurvCVM}
\fl
\Ps (r,\tau_1,\tau_2,D,s,\sigma_1,\ldots,\sigma_5, \chi, \kappa_1,\dots,\kappa_8;t) 
= - \frac{1}{2}\ave{\exp{-\int\plaind^d\xvec \rho(\xvec,t)}\tilderho(\xvec_0,t_0)} \\
\quad= - \frac{1}{2}\sum_{n=0}^\infty \frac{(-1)^n}{n!} 
\ave{\left(\int\plaind^d\xvec \rho(\xvec,t)\right)^n\tilderho(\xvec_0,t_0)}\nonumber\\
\quad\hat{=} - \frac{1}{2}\sum_{n=0}^\infty \frac{(-1)^n}{n!}\quad
\tikz[baseline=-2.5pt]{
\begin{scope}[rotate=-30]
  \draw [decorate,decoration={brace,amplitude=5pt}] (-128:1.4cm) -- (-172:1.4) node[pos=0.5,left,xshift=-0.07cm,yshift=0.14cm] {$n$};
  \draw[density] (-130:0.3) -- (-130:1.3);
  \path [postaction={decorate,decoration={raise=0ex,text along path,
  text={|\large|....}}}]
  (-155:1.2cm) arc (-155:-130:1.2cm);
  \draw[density] (-155:0.3) -- (-155:1.3);
  \draw[density] (-170:0.3) -- (-170:1.3);
\end{scope}
\draw[thick,fill=white,pattern=north east lines] (0,0) circle (0.3cm);
  \draw[density] (0:0.3) -- (0:0.8);
} 
\end{eqnarray}
\endnumparts
where the sum contains all binary tree diagrams with one incoming $\tilderho$ field and $n$ outgoing $\rho$ fields.

To account for the transmutations between the $\phi$ and $\psi$ fields in the sum in \Eqref{psurvCVM}, the density
field $\rho$ alone is not enough. To make use of the symmetries that will later allow for simplifications, I
consider the \emph{polarity} fields $\nu=\phi-\psi$ and $\tildenu=(\tau_1\tildephi-\tau_2\tildepsi)/(\tau_1+\tau_2)$.
The fields $\phi$, $\tildephi$, $\psi$, $\tildepsi$ are then mapped to the fields $\rho$, $\tilderho$, $\nu$, $\tildenu$, 
by means of linear combinations.
As  will become clearer, this a particularly simple mapping that allows the loop correction analysis.
The propagators $\ave{\nu\tildenu}$, $\ave{\rho\tildenu}$ and $\ave{\nu\tilderho}$
are stated explicitly in  \ref{sec:prop3}. Similarly, the couplings involving  $\rho$, $\tilderho$, $\nu$, $\tildenu$, 
are weighted linear combinations of $s$, $\sigma_1$, \ldots, $\sigma_5$, $ \chi$, $\kappa_1$, \dots, $\kappa_8$
and all observables may be parametrised in terms of these linear combinations weighted by powers of 
$\tau_1/(\tau_1+\tau_2)$ and $\tau_2/(\tau_1+\tau_2)$ as they enter $\tildenu$.
 \Eqref{psurvCVM} is thus simply a reparametrisation of the survival probability in terms of linear
combinations of couplings. Consider, for example, the branching non-linearity involving only $\rho$ and $\tilderho$ fields
\begin{equation}
\tikz[baseline=-2.5pt]{
\draw[density] (0.5,0) -- (0,0) node[at end,above=3pt,xshift=3pt] {${\zeta}$};
\draw[density] (130:0.5) -- (0,0);
\draw[density] (-130:0.5) -- (0,0);
}
\end{equation}
with ${\zeta}=\tau_2 \big( s\tau_2 +\tau_1 (\sigma_1+\sigma_2)\big)/\left( 2(\tau_1+\tau_2)^2\right)$,
and the four-legged non-linearity involving only $\rho$ and $\tilderho$ fields
\begin{equation}
\tikz[baseline=-2.5pt]{
\node [above=3pt] {${\lambda}$};
\draw[density] (-140:0.5) -- (40:0.5);
\draw[density] (140:0.5) -- (-40:0.5);
}
\end{equation}
with ${\lambda}=\left(\tau_2^2(\chi+\kappa_1)+\tau_1^2\kappa_2\right)/\left( 4(\tau_1+\tau_2)^2\right)$.

For example, for $n=3$, the sum in \Eqref{psurvCVM} has contributions from diagrams such as
 \begin{equation}
\tikz[baseline=-2.5pt]{
\draw[density] (0.5,0) -- (0,0) node [at end,above] {$\zeta$};
\draw[density] (140:0.8) -- (0,0);
\draw[density] (-140:0.8) -- (0,0);
\draw[density] (140:0.4)+(-140:0.4) -- (140:0.4) node [at end,above] {$\zeta$};}
\, , \quad
\tikz[baseline=-2.5pt]{
\draw[density] (0.5,0) -- (0,0);
\draw[density] (140:0.8) -- (140:0.4);
\draw[polarity] (140:0.4) -- (0,0);
\draw[density] (-140:0.8) -- (0,0);
\draw[density] (140:0.4)+(-140:0.4) -- (140:0.4);}
\, , \quad
\tikz[baseline=-2.5pt]{
\draw[density] (0.5,0) -- (0.25,0);
\draw[polarity] (0.25,0) -- (0,0);
\draw[density] (140:0.8) -- (140:0.2);
\draw[polarity] (140:0.2) -- (0,0);
\draw[density] (-140:0.8) -- (-140:0.4);
\draw[polarity] (-140:0.4) -- (0,0);
\draw[density] (140:0.4)+(-140:0.4) -- (140:0.4);}
\, ,
\end{equation}
where \tikz[baseline=-2.5pt]{ \draw[polarity] (0.8,0) -- (0,0) ;} represents the propagator $\ave{\nu\tildenu}$.

In what follows I show that any addend in \Eqref{psurvCVM} containing a $\nu$ or a $\tildenu$ field is asymptotically
 sub-leading and, therefore, its contribution can be neglected. From the propagators in 
  \ref{sec:prop3}, 
it follows that the propagators in real time, integrated over all space, are
\numparts
\begin{eqnarray}
\int \plaind^d\xvec \ave{ \nu(\xvec,t)  \tildenu(\xvec_0,t_0)}  =& -\exp{-(\tau_1+\tau_2+r)t} \, ,\\
\int \plaind^d\xvec \ave{ \rho(\xvec,t)  \tildenu(\xvec_0,t_0)}  =& 
\frac{\tau_2-\tau_1}{\tau_1+\tau_2}\exp{-(\tau_1+\tau_2+r)t} \, ,\\
\int \plaind^d\xvec \ave{ \nu(\xvec,t)  \tilderho(\xvec_0,t_0)}  =& 0 \, ,
\end{eqnarray}
\endnumparts
which decay exponentially faster in time $t$ than 
$\int \plaind^d\xvec \ave{ \rho(\xvec,t)  \tilderho(\xvec_0,t_0)}=2 e^{-rt}$ in \Eqref{CVM_nt}, since $r\to0$ while
$\tau_1, \tau_2>0$. Hence,
only binary tree diagrams containing $\rho$ and $\tilderho$ asymptotically contribute  to the sum in \Eqref{psurvCVM},
which gives
\begin{equation}
\fl\quad\Ps (r,\tau_1,\tau_2,D,s,\sigma_1,\ldots,\sigma_5, \chi, \kappa_1,\dots,\kappa_8;t) 
\simeq \frac{\exp{-rt}}{2\left(1+\frac{{\zeta}}{r}\left(1-\exp{-rt}\right)\right)} \, .
\end{equation}

Further, the fields $\nu$ and $\tildenu$ are also present in the renormalisation of the  non-linearities.
For example, ${\zeta}$ is renormalised diagrammatically by
\begin{equation}
\elabel{rho_branching_vertex}
\fl\quad\tikz[baseline=-2.5pt]{
\draw[black,fill=black] (0,0) circle (0.1);
\draw[density] (0.5,0) -- (0,0) node[at end,above,xshift=3pt] {${{\zeta}}_\renorm$};
\draw[density] (130:0.5) -- (0,0);
\draw[density] (-130:0.5) -- (0,0);
} \, 
=
\tikz[baseline=-2.5pt]{
\draw[density] (0.5,0) -- (0,0) node[at end,above,xshift=3pt] {${\zeta}$};
\draw[density] (130:0.5) -- (0,0);
\draw[density] (-130:0.5) -- (0,0);
} \, 
+
\tikz[baseline=-2.5pt]{
\draw[density] (0,0) circle (0.4);
\draw[density] (0.8,0) -- (0.4,0) node[at end,above=3pt,xshift=4pt] {$\zeta$};
\node (-0.5,0) [above=3pt,xshift=-14pt] {${\lambda}$};
\draw[density] ($(140:0.4)+(-0.4,0)$) -- (-0.4,0);
\draw[density] ($(-140:0.4)+(-0.4,0)$) -- (-0.4,0);
} \, 
+\tikz[baseline=-2.5pt]{
\draw[density] (0.4,0) arc (0:180:0.4);
\draw[polarity] (0.4,0) arc (0:-180:0.4);
\draw[density] (0.8,0) -- (0.4,0) ;
\draw[density] ($(140:0.4)+(-0.4,0)$) -- (-0.4,0);
\draw[density] ($(-140:0.4)+(-0.4,0)$) -- (-0.4,0);
} \,
+
\tikz[baseline=-2.5pt]{
\draw[polarity] (0,0) circle (0.4);
\draw[density] (0.8,0) -- (0.4,0) ;
\draw[density] ($(140:0.4)+(-0.4,0)$) -- (-0.4,0);
\draw[density] ($(-140:0.4)+(-0.4,0)$) -- (-0.4,0);
} \, 
+
\tikz[baseline=-2.5pt]{
\draw[density] (-0.4,0) arc (180:450:0.4);
\draw[polarity] (0,0.4) arc (90:180:0.4);
\draw[density] (0.8,0) -- (0.4,0) ;
\draw[density] ($(140:0.4)+(-0.4,0)$) -- (-0.4,0);
\draw[density] ($(-140:0.4)+(-0.4,0)$) -- (-0.4,0);
} \, 
+\ldots
\end{equation}
and similarly for the four-legged diagrams, which are, in fact, the governing non-linearities.
However, the choice of mapping between fields suitably fits the underlying symmetries of the original loops,
\Esref{loop_I1}--\eref{eq:CVM_I3}, and one can show that, crucially, 
 any loops containing $\nu$ or $\tildenu$ are ultraviolet convergent as far as the
phase transition at $r\to0$ is concerned. 
Expressing the loops formed with $\rho$, $\tilderho$, $\nu$, $\tildenu$ as linear combinations of the loops formed 
with $\phi$, $\tildephi$, $\psi$, $\tildepsi$ in Sec.~\ref{app_loop_integrals}, and assuming $\tau_1+\tau_2>0$, it follows
that the only ultraviolet divergence to be taken into account is
\numparts
\begin{eqnarray}
\fl
\mK &= 
\tikz[baseline=-2.5pt]{
\draw[density] (0,0) circle (0.4);
\draw[density] (0.5,0) -- (0.4,0);
\draw[density] (-0.5,0) -- (-0.4,0);
}\\
\fl
 &\corresponding \int\dbar^d\kvec_1\dbar^d\kvec_2\dbar^d\kvec_3\dbar^d\kvec_4
 \dbar\omega_1\dbar\omega_2\dbar\omega_3\dbar\omega_4
\, \deltabar(\kvec_2+\kvec_4)
\deltabar(\kvec_1+\kvec_3) \nonumber\\
\fl
&\quad\times
\deltabar(\omega_2+\omega_4)
\deltabar(\omega_1+\omega_3) 
\ave{\rho(\kvec_2,\omega_2)\tilderho(\kvec_1,\omega_1)}
\ave{\rho(\kvec_4,\omega_4)\tilderho(\kvec_3,\omega_3)} \nonumber\\
\fl
&= {2}{} \left(\frac{\tau_1+\tau_2}{\tau_2 4\pi D}\right)^{\frac{d}{2}}
  {{r}}^{\frac{d}{2}-1}   \Gamma\left(1-\frac{d}{2}\right)  \, ,
\end{eqnarray}
\endnumparts
whereas all the other loop integrals are ultraviolet convergent,
\begin{equation}
\fl
0 \, \hat{=}\, 
\tikz[baseline=-2.5pt]{
\draw[Bpolarity] (0.5,0) -- (0.4,0);
\draw[density] (-0.5,0) -- (-0.4,0);
\draw[density] (0.4,0) arc (0:180:0.4);
\draw[polarity] (0.4,0) arc (0:-180:0.4);
}
=
\tikz[baseline=-2.5pt]{
\draw[polarity] (0,0) circle (0.4);
\draw[Bpolarity] (0.5,0) -- (0.4,0);
\draw[Bpolarity] (-0.5,0) -- (-0.4,0);
}
=
\tikz[baseline=-2.5pt]{
\draw[density] (0,0.4) arc (90:360:0.4);
\draw[polarity] (0.4,0.0) arc (0:90:0.4);
\draw[density] (0.5,0) -- (0.4,0);
\draw[density] (-0.5,0) -- (-0.4,0);
}
=
\tikz[baseline=-2.5pt]{
\draw[density] (-0.4,0) arc (180:450:0.4);
\draw[polarity] (0,0.4) arc (90:180:0.4);
\draw[density] (0.5,0) -- (0.4,0);
\draw[density] (-0.5,0) -- (-0.4,0);
}
=
\tikz[baseline=-2.5pt]{
\draw[density] (0,0.4) arc (90:180:0.4);
\draw[polarity] (-0.4,0.0) arc (180:450:0.4);
\draw[Bpolarity] (0.5,0) -- (0.4,0);
\draw[Bpolarity] (-0.5,0) -- (-0.4,0);
}
=
\tikz[baseline=-2.5pt]{
\draw[density] (0.4,0) arc (360:450:0.4);
\draw[polarity] (0,0.4) arc (90:360:0.4);
\draw[Bpolarity] (0.5,0) -- (0.4,0);
\draw[Bpolarity] (-0.5,0) -- (-0.4,0);
}
=
\tikz[baseline=-2.5pt]{
\draw[density] (0,0.4) arc (90:270:0.4);
\draw[polarity] (0,-0.4) arc (270:450:0.4);
\draw[Bpolarity] (0.5,0) -- (0.4,0);
\draw[density] (-0.5,0) -- (-0.4,0);
}
=
\tikz[baseline=-2.5pt]{
\draw[polarity] (0,0.4) arc (90:270:0.4);
\draw[density] (0,-0.4) arc (270:450:0.4);
\draw[density] (0.5,0) -- (0.4,0);
\draw[Bpolarity] (-0.5,0) -- (-0.4,0);
}
=
\tikz[baseline=-2.5pt]{
\draw[density] (0,0.4) arc (90:180:0.4);
\draw[polarity] (0.4,0) arc (0:90:0.4);
\draw[density] (0,-0.4) arc (-90:0:0.4);
\draw[polarity] (-0.4,0) arc (180:270:0.4);
\draw[Bpolarity] (0.5,0) -- (0.4,0);
\draw[density] (-0.5,0) -- (-0.4,0);
} \, .
\end{equation}
In conclusion, the renormalisation of the reparametrised non-linearities is solely run by ${\lambda}$,
in the same way that, in the VM, $\chi$ is the governing non-linearity. It follows that the arguments in Sec.~\ref{epxdeltaVM}
for the VM apply equally to the CVM, producing the same result for the critical exponent $\delta$ in the CVM as in the VM.

\section{Discussion and conclusions \label{Conclus}}

In summary, the CVM has the same asymptotic behaviour as the VM at criticality and, therefore, the CVM belongs
to the VM universality class. The addition of the inner layer and the new interactions
with the external layer introduce changes in the short-term dynamics of the process, where the inner layer 
acts as an internal memory of previous states.
However, the new interactions are not strong enough to significantly change the long-range and
long-term behaviour of the process.

Yet, the interactions in the CVM manage to cause havoc to the renormalisation calculation and require linear 
combinations of fields and couplings to be considered.
The field theory encounters a range of interesting technical challenges. For example, since the VM at criticality is a particular
case of the CVM, the loop integrals involved in the CVM present multiple infrared divergencies.
Moreover, the renormalisation of the field theory in $\phi$ and $\psi$ involves ten loop integrals for a set of fifteen
coupled non-linearities. Nevertheless, the renormalisation scheme greatly simplifies by reparametrising
the non-linearities by suitable linear transformations of the fields that make use of the intrinsic
symmetries of the field theory of the CVM.

\ack
I am indebted to Gunnar Pruessner, who has taught me almost all I know about field theory.
I would also like to thank Michael Gastner and Be\'ata Oborny for fruitful discussion about the consensus time of the
CVM. I am grateful to Luca Cocconi, Miguel Mu\~{n}oz, Mauro Mobilia, and the Non-Equilibrium 
Systems group at Imperial College for useful discussions.

\appendix
\section{Ward identity}
\label{Ward}
In this section I adapt the derivation of the Ward identity in \cite{bordeu2019volume} to the Voter Model.
The action of the VM without irrelevant terms is, from \Esref{VMmA}
and \eref{eq:VMmA1},
\begin{equation}
\fl\quad
\mA\left([\phi,\tildephi];D,r,s,\chi \right) = \int \plaind^d\xvec\plaind t \Big\{ 
\tildephi (\partial_t -D \nabla^2 + r) \phi
- {s}\tildephi^2\phi + {\chi} \tildephi^2\phi^2 
\Big\} \,.
\end{equation}
The Ward identity, \Eqref{W_ID}, has its origin in a symmetry of the action when shifting the field $\phi(x,t)$ by a constant 
$\Sigma$,
\begin{equation}
\fl\quad\mA\left([\phi+\Sigma,\tildephi];D,r,s,\chi \right) = \mA\left([\phi,\tildephi];D,r,s-2\chi\Sigma,\chi \right) 
+ \int \plaind^d\xvec\plaind t \Big\{ r\Sigma\tildephi -s\Sigma\tildephi^2+\chi\Sigma^2\tildephi^2\Big\} \,.
\elabel{actionshift}
\end{equation}
To simplify the notation, let
\begin{eqnarray}
\mA &= \mA\left([\phi,\tildephi];D,r,s,\chi \right)\\
\mA' &= \mA\left([\phi+\Sigma,\tildephi];D,r,s,\chi \right)\\
\mA'' &= \mA\left([\phi,\tildephi];D,r,s-2\chi\Sigma,\chi \right)
\end{eqnarray}
so that \Eqref{actionshift} reads
\begin{equation}
\mA' = \mA''+ \int \plaind^d\xvec\plaind t \Big\{ r\Sigma\tildephi -s\Sigma\tildephi^2+\chi\Sigma^2\tildephi^2\Big\} \,,
\end{equation}
and define the path integral
\begin{equation}
\ave{\mO}_\mA = \int \mD [\phi,\tildephi] \, \mO \, \exp{-\mA},
\end{equation}
and equivalently for the actions $\mA'$ and $\mA''$, for any observable $\mO$. To derive the Ward identity,
consider the observable $\ave{\phi(\xvec_2,t_2)\phi(\xvec_1,t_1)\tildephi(\xvec_0,t_0)}_\mA $. Since $\phi$ is
a dummy variable in the path integral, the integral is invariant under a shift of the field by a constant $\Sigma$,
\begin{eqnarray}
\elabel{obs_shift}
\fl \quad\ave{\phi(\xvec_2,t_2)\phi(\xvec_1,t_1)\tildephi(\xvec_0,t_0)}_\mA 
=  \ave{\Big(\phi(\xvec_2,t_2)+\Sigma\Big)\Big(\phi(\xvec_1,t_1)+\Sigma\Big)\tildephi(\xvec_0,t_0)}_{\mA'} \nonumber\\
=  \ave{\phi(\xvec_2,t_2)\phi(\xvec_1,t_1)\tildephi(\xvec_0,t_0)}_{\mA'}
+\Sigma\ave{\phi(\xvec_1,t_1)\tildephi(\xvec_0,t_0)}_{\mA'} \\
\quad +\Sigma\ave{\phi(\xvec_2,t_2)\tildephi(\xvec_0,t_0)}_{\mA'}
+\Sigma^2\ave{\tildephi(\xvec_0,t_0)}_{\mA'} \nonumber \,.
\end{eqnarray}
Differentiating \Eqref{obs_shift} with respect to $\Sigma$ and evaluating at $\Sigma=0$ gives
\begin{eqnarray}
\elabel{Ward_expanded}
 0 =& 2\chi\partial_s 
 \ave{\phi(\xvec_2,t_2)\phi(\xvec_1,t_1)\tildephi(\xvec_0,t_0)}_{\mA} \\&
 - r \int \plaind^d\xvec'\plaind t' \ave{\phi(\xvec_2,t_2)\phi(\xvec_1,t_1)\tildephi(\xvec_0,t_0)\tildephi(\xvec',t')}_{\mA}
\nonumber \\&
 +s \int \plaind^d\xvec'\plaind t' \ave{\phi(\xvec_2,t_2)\phi(\xvec_1,t_1)\tildephi(\xvec_0,t_0)\tildephi^2(\xvec',t')}_{\mA}
\nonumber \\&
 +\ave{\phi(\xvec_1,t_1)\tildephi(\xvec_0,t_0)}_{\mA}
+\ave{\phi(\xvec_2,t_2)\tildephi(\xvec_0,t_0)}_{\mA} \nonumber \,,
\end{eqnarray}
since $\mA'=\mA$ at $\Sigma=0$. The path integral in the first term of the right-hand side of \Eqref{Ward_expanded} is 
\begin{eqnarray}
\elabel{Ward_term1}
\fl\ave{\phi(\xvec_2,t_2)\phi(\xvec_1,t_1)\tildephi(\xvec_0,t_0)}_{\mA} 
\corresponding\quad 2\,
\tikz[baseline=-2.5pt]{
\draw[Aactivity] (0.5,0) -- (0,0) node[at start, above] {\scriptsize $0$};
\draw[Aactivity] (130:0.5) -- (0,0) node[at start, left] {\scriptsize $1$};
\draw[Aactivity] (-130:0.5) -- (0,0) node[at start, left] {\scriptsize $2$};
\draw[black,fill=black] (0,0) circle (0.1);
}
\\\fl
\quad = 2 s_{\textnormal{\scriptsize eff}}
\int \dbar^d\kvec_0 \dbar^d\kvec_0' \dbar^d\kvec_1 \dbar^d\kvec_1' \dbar^d\kvec_2\dbar^d\kvec_2'
\dbar\omega_0\dbar\omega_0' \dbar\omega_1\dbar\omega_1' \dbar\omega_2 \dbar\omega_2' \nonumber\\\fl
\quad\quad\times \deltabar(\omega_0'+\omega_1'+\omega_2')\deltabar(\kvec_0'+\kvec_1'+\kvec_2') \,
\exp{-\imag(\omega_0t_0+\omega_1t_1+\omega_2t_2)
+\imag(\kvec_0\xvec_0+\kvec_1\xvec_1+\kvec_2\xvec_2)}\nonumber\\\fl
\quad\quad\times\ave{\phi(\omega_0',\kvec_0')\tildephi(\omega_0,\kvec_0)}
\ave{\phi(\omega_1,\kvec_1)\tildephi(\omega_1',\kvec_1')}
\ave{\phi(\omega_2,\kvec_2)\tildephi(\omega_2',\kvec_2')}\nonumber \\\fl
\quad = 2 s_{\textnormal{\scriptsize eff}}\int \dbar^d\kvec_1\dbar^d\kvec_2 \dbar\omega_1\dbar\omega_2 \,
\exp{-\imag(\omega_1(t_1-t_0)+\omega_2(t_2-t_0))
+\imag(\kvec_1(\xvec_1-\xvec_0)+\kvec_2(\xvec_2-\xvec_0))}\nonumber\\\fl
\quad\quad\times
\ave{\phi(\omega_1+\omega_2,\kvec_1+\kvec_2)\tildephi(-\omega_1-\omega_2,-\kvec_1-\kvec_2)}\nonumber\\\fl
\quad\quad\times
\ave{\phi(\omega_1,\kvec_1)\tildephi(-\omega_1,-\kvec_1)}
\ave{\phi(\omega_2,\kvec_2)\tildephi(-\omega_2,-\kvec_2)}\nonumber
\,.
\end{eqnarray}
The second path integral in \Eqref{Ward_expanded} has three contributions,
\begin{eqnarray}
\elabel{Ward_term2}
\fl \int \plaind^d\xvec'\plaind t'\ave{\phi(\xvec_2,t_2)\phi(\xvec_1,t_1)\tildephi(\xvec_0,t_0)\tildephi(\xvec',t')}_{\mA}\\\fl\quad
\corresponding \int \plaind^d\xvec'\plaind t' \Big\{
\tikz[baseline=-2.5pt]{\draw[Aactivity] (-0.5,0) -- (0.5,0) node[at start, above] {\scriptsize $1$} node[at end, above] {\scriptsize $0$};}
\times
\tikz[baseline=-2.5pt]{\draw[Aactivity] (-0.5,0) -- (0.5,0) node[at start, above] {\scriptsize $2$} node[at end, above] {\scriptsize $'$};}
+
\tikz[baseline=-2.5pt]{\draw[Aactivity] (-0.5,0) -- (0.5,0) node[at start, above] {\scriptsize $2$} node[at end, above] {\scriptsize $0$};}
\times
\tikz[baseline=-2.5pt]{\draw[Aactivity] (-0.5,0) -- (0.5,0) node[at start, above] {\scriptsize $1$} node[at end, above] {\scriptsize $'$};}
+ 4
\tikz[baseline=-2.5pt]{
\draw[Aactivity] (140:0.5) -- (-40:0.5) node[at start, above] {\scriptsize $1$} node[at end, below] {\scriptsize $'$};
\draw[Aactivity] (-140:0.5) -- (40:0.5) node[at start, below] {\scriptsize $2$} node[at end, above] {\scriptsize $0$};
\draw[black,fill=black] (0,0) circle (0.1);
}\Big\} 
\nonumber
\\\fl\quad
=
\frac{1}{r}\ave{\phi(\xvec_1,t_1)\tildephi(\xvec_0,t_0)}
+\frac{1}{r}\ave{\phi(\xvec_2,t_2)\tildephi(\xvec_0,t_0)}\nonumber\\\fl\quad
\quad+\frac{4\chi_{\textnormal{\scriptsize eff}}}{r}\int \dbar^d\kvec_1\dbar^d\kvec_2 \dbar\omega_1\dbar\omega_2 \,
\exp{-\imag(\omega_1(t_1-t_0)+\omega_2(t_2-t_0))
+\imag(\kvec_1(\xvec_1-\xvec_0)+\kvec_2(\xvec_2-\xvec_0))}\nonumber\\\fl
\quad\quad\times
\ave{\phi(\omega_1+\omega_2,\kvec_1+\kvec_2)\tildephi(-\omega_1-\omega_2,-\kvec_1-\kvec_2)}\nonumber\\\fl
\quad\quad\times
\ave{\phi(\omega_1,\kvec_1)\tildephi(-\omega_1,-\kvec_1)}
\ave{\phi(\omega_2,\kvec_2)\tildephi(-\omega_2,-\kvec_2)}\nonumber
\,,
\end{eqnarray}
where integrating over space and time amounts to evaluating the propagator in Fourier space at 
$\omega'=0$ and $\kvec'=\gpvec{0}$,
\begin{equation}
\fl\quad
\int \plaind^d\xvec'\plaind t'\ave{\phi(\xvec,t)\tildephi(\xvec',t')}=\int\dbar^d\kvec\dbar\omega
\ave{\phi(\omega,\kvec)\tildephi(\omega'=0,\kvec'=\gpvec{0})} = \frac{1}{r} \,.
\end{equation}
The third term in \Eqref{Ward_expanded}, with coupling $s$, vanishes
because the action does not provide any vertex that allows to pair each creator field with an annihilator field.
Substituting \Esref{Ward_term1} and \eref{eq:Ward_term2} in \Eqref{Ward_expanded}, and dividing 
out the common integral, which amounts to amputating diagrams,
 gives the Ward identity, 
\begin{equation}
0 = \chi\partial_s s_{\textnormal{\scriptsize eff}} - \chi_{\textnormal{\scriptsize eff}}\, ,
\elabel{Ward_ID}
\end{equation} 
which is a generalisation of the 
VM Lagrangian invariance in \cite{Munoz:1997}.

The Ward identity in \Eqref{Ward_ID} also holds in the diffusion-limited pair annihilation process field theory
 \cite{peliti1986renormalisation} by virtue of the rapidity reversal of the VM field theory, \Eqref{rap_rev}.
 In this case, the origin of the Ward identity lies in the symmetry of the action under a shift of the creation field by 
 a constant (instead of the annihilation field).

\section{Propagators involving $\phi$ and $\psi$}
\label{app_propsCVM1}
From \Eqref{CVM_A0}, the four  response propagators $\ave{\phi\tildephi}$, $\ave{\psi\tildepsi}$,
$\ave{\phi\tildepsi}$, $\ave{\psi\tildephi}$, are read off from the inverse of the bilinear interaction matrix as follows:
\begin{eqnarray}
\elabel{eqA1}
\tikz[baseline=-2.5pt]{
\node at (0.5,0) [above] {$\tildephi$};
\node at (-0.5,0) [above] {$\phi$};
\draw[Aactivity] (0.5,0) -- (-0.5,0) node[at end,above] {};
}
&=
\, \tikz[baseline=-2.5pt]{ \draw[Aactivity] (0.4,0) -- (-0.4,0) ; }
\,+
\tikz[baseline=-2.5pt]{
\draw[Aactivity] (0.5,0) -- (0.25,0) ; 
\draw[substrate] (0.25,0) -- (-0.25,0) ; 
\draw[Aactivity] (-0.25,0) -- (-0.5,0) ; 
}\,+
\tikz[baseline=-2.5pt]{
\draw[Aactivity] (1.25,0) -- (1,0) ; 
\draw[substrate] (1,0) -- (0.5,0) ; 
\draw[Aactivity] (0.5,0) -- (0.25,0) ; 
\draw[substrate] (0.25,0) -- (-0.25,0) ; 
\draw[Aactivity] (-0.25,0) -- (-0.5,0) ; 
}\,+\ldots \nonumber
\\&\hat{=}\ave{\phi(\kvec,\omega)\tildephi(\kvec',\omega')}
\\&=
\frac{\deltabar(\omega+\omega')\deltabar(\kvec+\kvec') \left(-\imag \omega  +r_2\right)}{\left(-\imag \omega  + D \kvec^2 + r_1\right)\left(-\imag \omega  +r_2\right)-\tau_1\tau_2} \, , \nonumber \\
\tikz[baseline=-2.5pt]{
\node at (0.5,0) [above] {$\tildepsi$};
\node at (-0.5,0) [above] {$\psi$};
\draw[substrate] (0.5,0) -- (-0.5,0) node[at end,above] {};
}\, &=
\tikz[baseline=-2.5pt]{ \draw[substrate] (0.4,0) -- (-0.4,0) ; }
\,+
\tikz[baseline=-2.5pt]{
\draw[substrate] (0.5,0) -- (0.125,0) ; 
\draw[Aactivity] (0.125,0) -- (-0.125,0) ; 
\draw[substrate] (-0.125,0) -- (-0.5,0) ; 
}\,+
\tikz[baseline=-2.5pt]{
\draw[substrate] (1.125,0) -- (0.75,0) ; 
\draw[Aactivity] (0.75,0) -- (0.5,0) ; 
\draw[substrate] (0.5,0) -- (0.125,0) ; 
\draw[Aactivity] (0.125,0) -- (-0.125,0) ; 
\draw[substrate] (-0.125,0) -- (-0.5,0) ; 
}\,+\ldots \nonumber
\\&\hat{=}\ave{\psi(\kvec,\omega)\tildepsi(\kvec',\omega')} \\
&=
\frac{\deltabar(\omega+\omega')\deltabar(\kvec+\kvec') \left( -\imag \omega  + D \kvec^2 + r_1 \right)}{\left(-\imag \omega  + D \kvec^2 + r_1\right)\left(-\imag \omega  +r_2\right)-\tau_1\tau_2} \, , \nonumber \\
\tikz[baseline=-2.5pt]{
\node at (0.5,0) [above] {$\tildepsi$};
\node at (-0.5,0) [above] {$\phi$};
\draw[Aactivity] (0,0) -- (-0.5,0) ;
\draw[substrate] (0.5,0) -- (0,0) ;
}\,&=
\tikz[baseline=-2.5pt]{
\draw[Aactivity] (-0.4,0) -- (0,0) ;
\draw[substrate] (0,0) -- (0.4,0) ;
 }
\,+
\tikz[baseline=-2.5pt]{
\draw[Aactivity] (-0.625,0) -- (-0.375,0) ;
\draw[substrate] (-0.375,0) -- (0,0) ;
\draw[Aactivity] (0,0) -- (0.25,0) ;
\draw[substrate] (0.25,0) -- (0.625,0) ;
}\,+
\tikz[baseline=-2.5pt]{
\draw[Aactivity] (-0.7,0) -- (-0.55,0) ;
\draw[substrate] (-0.55,0) -- (-0.35,0) ;
\draw[Aactivity] (-0.35,0) -- (-0.2,0) ;
\draw[substrate] (-0.2,0) -- (0,0) ;
\draw[substrate] (0.7,0) -- (0.5,0) ;
\draw[Aactivity] (0.5,0) -- (0.35,0) ;
\draw[substrate] (0.35,0) -- (0.15,0) ;
\draw[Aactivity] (0.15,0) -- (0,0) ;
}\,+\ldots \nonumber
\\&
\hat{=}\ave{\phi(\kvec,\omega)\tildepsi(\kvec',\omega')}\\
&=\frac{\deltabar(\omega+\omega')\deltabar(\kvec+\kvec') \,\tau_1}{\left(-\imag \omega  + D \kvec^2 + r_1\right)\left(-\imag \omega  +r_2\right)-\tau_1\tau_2} \, , \nonumber \\
\elabel{eqA4}
\tikz[baseline=-2.5pt]{
\node at (0.5,0) [above] {$\tildephi$};
\node at (-0.5,0) [above] {$\psi$};
\draw[Aactivity] (0.5,0) -- (0,0) ;
\draw[substrate] (0,0) -- (-0.5,0) ;
}\,&=
\tikz[baseline=-2.5pt]{
\draw[Aactivity] (0.4,0) -- (0,0) ;
\draw[substrate] (0,0) -- (-0.4,0) ;
 }
\,+
\tikz[baseline=-2.5pt]{
\draw[Aactivity] (0.625,0) -- (0.375,0) ;
\draw[substrate] (0.375,0) -- (0,0) ;
\draw[Aactivity] (0,0) -- (-0.25,0) ;
\draw[substrate] (-0.25,0) -- (-0.625,0) ;
}\,+
\tikz[baseline=-2.5pt]{
\draw[substrate] (-0.7,0) -- (-0.5,0) ;
\draw[Aactivity] (-0.5,0) -- (-0.35,0) ;
\draw[substrate] (-0.35,0) -- (-0.15,0) ;
\draw[Aactivity] (-0.15,0) -- (0,0) ;
\draw[Aactivity] (0.7,0) -- (0.55,0) ;
\draw[substrate] (0.55,0) -- (0.35,0) ;
\draw[Aactivity] (0.35,0) -- (0.2,0) ;
\draw[substrate] (0.2,0) -- (0,0) ;
}\,+\ldots \nonumber
\\&
\hat{=}\ave{\psi(\kvec,\omega)\tildephi(\kvec',\omega')}\\
&=\frac{\deltabar(\omega+\omega')\deltabar(\kvec+\kvec') \,\tau_2}{\left(-\imag \omega  + D \kvec^2 + r_1\right)\left(-\imag \omega  +r_2\right)-\tau_1\tau_2} \, . \nonumber
\end{eqnarray}

\section{Loop integrals with $\phi$ and $\psi$}
\label{app_loop_integrals}
Let $r_i=\tau_i+{r}$, $i\in\{1,2\}$, in \Eqref{eqA1}--\eref{eq:eqA4} and
\begin{equation}
\tildeU = \frac{(\tau_1+\tau_2)^{\frac{d}{2}-2}}{2\tau_2(4\pi D)^{d/2}}  \Gamma\left(1-\frac{d}{2}\right) \, ,
\end{equation}
then the loop integrals involving the fields $\phi$, $\tildephi$, $\psi$, $\tildepsi$,
are
\begin{eqnarray}
I_1 &{=}\,
\tikz[baseline=-2.5pt]{
\draw[Aactivity] (0,0) circle (0.4);
\draw[Aactivity] (0.5,0) -- (0.4,0);
\draw[Aactivity] (-0.5,0) -- (-0.4,0);
}
\,\hat{=} \tildeU \left( \tau_1\tau_2 + \tau_2^2 \left(\frac{{r}}{\tau_2}\right)^{\frac{d}{2}-1}\right)
\, ,\elabel{loop_I1}\\
I_{2} &{=}\,
\tikz[baseline=-2.5pt]{
\draw[Aactivity] (0,0.4) arc (90:360:0.4);
\draw[Bsubstrate] (0.4,0.0) arc (0:90:0.4);
\draw[Aactivity] (0.5,0) -- (0.4,0);
\draw[Aactivity] (-0.5,0) -- (-0.4,0);
}
\,\hat{=} \, \tildeU \left( -\tau_1\tau_2 + \tau_1\tau_2 \left(\frac{{r}}{\tau_2}\right)^{\frac{d}{2}-1}\right)
\, ,\\
I_{3} &{=} \,
\tikz[baseline=-2.5pt]{
\draw[Aactivity] (-0.4,0) arc (180:450:0.4);
\draw[Bsubstrate] (0,0.4) arc (90:180:0.4);
\draw[Aactivity] (0.5,0) -- (0.4,0);
\draw[Aactivity] (-0.5,0) -- (-0.4,0);
}
\,\hat{=} \, \tildeU \left( -\tau_2^2 + \tau_2^2 \left(\frac{{r}}{\tau_2}\right)^{\frac{d}{2}-1}\right)
\, ,\\
I_4 &{=}\,
\tikz[baseline=-2.5pt]{
\draw[Aactivity] (0.5,0) -- (0.4,0);
\draw[Aactivity] (-0.5,0) -- (-0.4,0);
\draw[Aactivity] (0.4,0) arc (0:180:0.4);
\draw[Bsubstrate] (0.4,0) arc (0:-180:0.4);
}
\,\hat{=} \, \tildeU \left( \tau_1\tau_2 +2\tau_2^2+ \tau_1\tau_2 \left(\frac{{r}}{\tau_2}\right)^{\frac{d}{2}-1}\right)
\, ,\\
I_{5} &{=}\,
\tikz[baseline=-2.5pt]{
\draw[Aactivity] (0,0.4) arc (90:270:0.4);
\draw[Bsubstrate] (0,-0.4) arc (270:450:0.4);
\draw[substrate] (0.5,0) -- (0.4,0);
\draw[Aactivity] (-0.5,0) -- (-0.4,0);
}
\,\hat{=} \, \tildeU \left( -{\tau_1^2}{} + {\tau_1^2}{} \left(\frac{{r}}{\tau_2}\right)^{\frac{d}{2}-1}\right)
\, ,\\
I_{6} &{=}\,
\tikz[baseline=-2.5pt]{
\draw[Aactivity] (0,0.4) arc (90:180:0.4);
\draw[Bsubstrate] (0.4,0) arc (0:90:0.4);
\draw[Aactivity] (0,-0.4) arc (-90:0:0.4);
\draw[Bsubstrate] (-0.4,0) arc (180:270:0.4);
\draw[substrate] (0.5,0) -- (0.4,0);
\draw[Aactivity] (-0.5,0) -- (-0.4,0);
}
\,\hat{=} \, I_{2}\, ,\\
I_{7} &{=}\,
\tikz[baseline=-2.5pt]{
\draw[Bsubstrate] (0,0.4) arc (90:270:0.4);
\draw[Aactivity] (0,-0.4) arc (270:450:0.4);
\draw[Aactivity] (0.5,0) -- (0.4,0);
\draw[substrate] (-0.5,0) -- (-0.4,0);
}
\,\hat{=} \, I_{3} 
\, ,\\
I_{8} &{=}\,
\tikz[baseline=-2.5pt]{
\draw[Aactivity] (0.4,0) arc (360:450:0.4);
\draw[Bsubstrate] (0,0.4) arc (90:360:0.4);
\draw[substrate] (0.5,0) -- (0.4,0);
\draw[substrate] (-0.5,0) -- (-0.4,0);
}
\,\hat{=} \, \tildeU \left( \tau_2^2 + \tau_1\tau_2 \left(\frac{{r}}{\tau_2}\right)^{\frac{d}{2}-1}\right)
\, ,\\
I_{9} &{=}\,
\tikz[baseline=-2.5pt]{
\draw[Aactivity] (0,0.4) arc (90:180:0.4);
\draw[Bsubstrate] (-0.4,0.0) arc (180:450:0.4);
\draw[substrate] (0.5,0) -- (0.4,0);
\draw[substrate] (-0.5,0) -- (-0.4,0);
}
\,\hat{=} \, \tildeU \left( \tau_1\tau_2 + {\tau_1^2} \left(\frac{{r}}{\tau_2}\right)^{\frac{d}{2}-1}\right)
\, ,\\
\elabel{CVM_I3}
I_{10} &{=}\,
\tikz[baseline=-2.5pt]{
\draw[Bsubstrate] (0,0) circle (0.4);
\draw[substrate] (0.5,0) -- (0.4,0);
\draw[substrate] (-0.5,0) -- (-0.4,0);
}
\,\hat{=} \,\frac{\Lambda^d}{2\tau_2} + I_9
\, ,
\end{eqnarray}
where $\Lambda<\infty$ is an ultraviolet cutoff to regularise the quadratic divergence of $I_{10}$ at $d=d_c$
 as $|\kvec|\to\infty$. The loop $I_{10}$ is UV finite with a quadratic divergence entering only as the internal layer becomes 
 inert in the absence of internalisation, $\tau_2\to0$.
 All loops above have the same divergence as $r\to0$, the critical point of the CVM, up to an amplitude in powers of $\tau_1$
 and $\tau_2$ that is effectively accounted for by the linear combination of fields introduced with $\tildenu$.
 In any loops that contain $\nu$ and $\tildenu$ fields, which are constructed by subtracting fields $\phi$, $\psi$ and 
 $\tildephi$, $\tildepsi$ respectively, UV divergencies will therefore by symmetry vanish.

\section{Uniqueness of the critical point of the CVM}
\label{app_criticalpoint}
{The critical point is found in the divergence of the loop integrals.
For example, the integral $I_1$ in \Eqref{loop_I1} 
is, assuming $r_1,r_2,\tau_1,\tau_2>0$,
\begin{equation}
\fl
I_1 = \frac{\Gamma\left(1-\frac{d}{2}\right)}{2(r_2^2+\tau_1\tau_2)(4\pi D)^{d/2}} \left[
\tau_1\tau_2(r_1+r_2)^{d/2-1} + r_2^2 \left( r_1-\frac{\tau_1\tau_2}{r_2}\right)^{d/2-1}
\right] \,.
\end{equation}
This indicates the presence of two divergencies: $r_1+r_2\to0$, which recovers the original VM,
 and $\bar\varepsilon=r_1-{\tau_1\tau_2}/{r_2}\to0$, which defines the  parameter regime of the CVM at criticality.
 Let the masses  $r_1$ and $r_2$ be parametrised by $r_{i} = \tau_{i} + \theta_{i}$ with $i\in\{1,2\}$.
  The distance to the critical point $\bar\varepsilon$ reads
 \begin{equation}
 \bar\varepsilon=\tau_{1} + \theta_{1}-\frac{\tau_1\tau_2}{\tau_{2} + \theta_{2}} 
 =\frac{\theta_1\tau_2+\theta_2\tau_1+\theta_1\theta_2}{\tau_{2} + \theta_{2}} \,.
 \end{equation}
Taking the limit $\theta_i\to0$ while keeping $\theta_j>0$, with $j\in\{1,2\}$ and $j\neq i$, gives
\begin{equation}
\lim_{\theta_i\to0} \bar\varepsilon = \frac{\theta_j\tau_i}{\tau_2+\theta_2} >0 \,.
\end{equation}
Since it is necessary to take both limits, ${\theta_1\to0}$ and ${\theta_2\to0}$,
 simultaneously in order to find the critical point,
\begin{equation}
\lim_{\theta_1\to0} \lim_{\theta_2\to0} \bar\varepsilon =0 \,,
\end{equation}
the critical point of the CVM is unique and therefore $r_i$ can be parametrised by $r_{i} = \tau_{i} + r$.
}

\section{Propagators involving $\rho$ and $\nu$}\label{sec:prop3}
The propagators $\ave{\nu\tildenu}$, $\ave{\rho\tildenu}$, $\ave{\nu\tilderho}$, are
\begin{eqnarray}
\fl \quad\quad\,
\tikz[baseline=-2.5pt]{
\node at (0.5,0) [above] {$\tildenu$};
\node at (-0.5,0) [above] {$\nu$};
\draw[polarity] (0.5,0) -- (-0.5,0) node[at end,above] {};
}\hat{=}\ave{\nu(\kvec,\omega)\tildenu(\kvec',\omega')}
\\
=
{\deltabar(\omega+\omega')\deltabar(\kvec+\kvec')}{}\nonumber
\frac{(\tau_1+\tau_2)(-\imag\omega)+\tau_2D\kvec^2+\tau_1r_2+\tau_2r_1-2\tau_1\tau_2}{\left(\tau_1+\tau_2\right) 
\left[\left(-\imag \omega  + D \kvec^2 + r_1\right)\left(-\imag \omega  +r_2\right)-\tau_1\tau_2\right]} \nonumber \, ,\\
\fl\quad\quad\,
\tikz[baseline=-2.5pt]{
\node at (0.5,0) [above] {$\tildenu$};
\node at (-0.5,0) [above] {$\rho$};
\draw[polarity] (0.5,0) -- (0,0) node[at end,above] {};
\draw[density] (0,0) -- (-0.5,0) node[at end,above] {};
}\hat{=}\ave{\rho(\kvec,\omega)\tildenu(\kvec',\omega')}\\
=
{\deltabar(\omega+\omega')\deltabar(\kvec+\kvec')}{}\nonumber 
\frac{\left(\tau_1-\tau_2\right)(-\imag \omega) -\tau_2 D \kvec^2 + \tau_1r_2 - \tau_2r_1}{\left(\tau_1+\tau_2\right)\left[\left(-\imag \omega  + D \kvec^2 + r_1\right)\left(-\imag \omega  +r_2\right)-\tau_1\tau_2\right]} \nonumber \, ,\\
\fl\quad\quad\,
\tikz[baseline=-2.5pt]{
\node at (0.5,0) [above] {$\tilderho$};
\node at (-0.5,0) [above] {$\nu$};
\draw[density] (0.5,0) -- (0,0) node[at end,above] {};
\draw[polarity] (0,0) -- (-0.5,0) node[at end,above] {};
}\hat{=}\ave{\nu(\kvec,\omega)\tilderho(\kvec',\omega')}\\
=
\deltabar(\omega+\omega')\deltabar(\kvec+\kvec')
\frac{
 - D \kvec^2 + r_2-r_1+\tau_1 - \tau_2}{\left(-\imag \omega  + D \kvec^2 + r_1\right)\left(-\imag \omega  +r_2\right)-\tau_1\tau_2} \nonumber \, .
\end{eqnarray}

\bibliography{books,articles}
\end{document}